\documentclass[paper,twoside]{JHEP3}
\newif\ifpdf\ifx\pdfoutput\undefined\pdffalse\else\pdfoutput=1\pdftrue\fi
 
\usepackage{epsfig,amsfonts,amsmath,array,mathrsfs,wick}
\newcommand{\pbs}[1]{\let\temp=\\#1\let\\=\temp}
\renewcommand{\theequation}{\thesection.\arabic{equation}}
\def\be{\begin{equation}}\def\ee{\end{equation}}
\def\cvp{\raise 2pt\hbox{,}}

\def\Trf{\mathop{\rm Tr_{F}}\nolimits}  \def\im{\mathop{\rm Im}\nolimits}
\def\re{\mathop{\rm Re}\nolimits} \def\diag{\mathop{\rm
diag}\nolimits} 

 \def\d{{\rm d}}\def\nn{{\cal
N}} 

\def\bigllangle{\bigl\langle\!\bigl\langle}
\def\bigrrangle{\bigr\rangle\!\bigr\rangle}
\def\Bigllangle{\Bigl\langle\!\!\Bigl\langle}
\def\Bigrrangle{\Bigr\rangle\!\!\Bigr\rangle} \def\Nf{N_{\mathrm f}}
\def\suN{{\rm SU}(N)} \def\uN{{\rm U}(N)} 
  
\def\F{\mathscr F}\def\Fb{\mathcal F} \def\Fp{\mathcal
F_{\mathrm{s}}}\def\Fd{\mathcal F_{\mathrm{d}}}

\def\wdv{W_{\text{DV}}}\def\wmm{W_{\text{MM}}}\def\wvy{W_{\text{VY}}}
\def\whdv{\hat{W}_{\text{DV}}} \def\R{\mathscr R}\def\S{\mathscr
S}\def\Y{\mathscr Y}
\def\wl{W_{\rm low}}\def\wt{W_{\rm tree}}

\def\La{\Lambda}

\def\u{{\rm U}(1)}

\def\plb#1#2#3{{\it Phys.\ Lett.\ }{\bf B #1} (#2) #3}
\def\npb#1#2#3{{\it Nucl.\ Phys.\ }{\bf B #1} (#2) #3}

\def\jhep#1#2#3{{\it J. High Energy Phys.\ }{\bf #1} (#2) #3}
\def\prd#1#2#3{{\it Phys.\ Rev.\ }{\bf D #1} (#2) #3}

\def\atmp#1#2#3{{\it Adv.\ Theor.\ Math.\ Phys.\ }{\bf #1} (#2) #3}

\def\jmp#1#2#3{{\it J.\ Math.\ Phys.\ }{\bf #1} (#2) #3}

\def\mpla#1#2#3{{\it Mod.\ Phys.\ Lett.\ }{\bf A #1} (#2) #3}

\title{The Proof of the Dijkgraaf-Vafa Conjecture
and application to the mass gap and confinement problems}
\author{Frank Ferrari\\
Service de Physique Th\'eorique et Math\'ematique\\
Universit\'e Libre de Bruxelles and International Solvay Institutes\\
Campus de la Plaine, CP 231, B-1050 Bruxelles, Belgique\\
E-mail: \email{frank.ferrari@ulb.ac.be}}

\abstract{Using generalized Konishi anomaly equations, it is known
that one can express, in a large class of supersymmetric gauge
theories, all the chiral operators expectation values in terms of a
finite number of a priori arbitrary constants. We show that these
constants are fully determined by the requirement of gauge invariance
and an additional anomaly equation. The constraints so obtained turn
out to be equivalent to the extremization of the Dijkgraaf-Vafa
quantum glueball superpotential, with all terms (including the
Veneziano-Yankielowicz part) unambiguously fixed. As an application,
we fill non-trivial gaps in existing derivations of the mass gap and
confinement properties in super Yang-Mills theories.}
\preprint{LPTENS-05/28, hep-th/0602249}
\begin{document}
\setcounter{footnote}{0}

%
\section{General presentation}
\setcounter{equation}{0}

The aim of this paper is to present a general proof of the
Dijkgraaf-Vafa quantum equations of motion \cite{DV},
\be\label{DVqem} \frac{\partial\wdv}{\partial s_{I}} = 0\, ,\ee
that determine the glueball superfields expectation values
$s_{I}=\langle S_{I}\rangle$ in a large class of $\nn=1$
supersymmetric gauge theories. The holomorphic function $\wdv(s_{I})$
is the sum of a suitable Veneziano-Yankielowicz term \cite{VY} and of
an infinite power series in the $s_{I}$s. The equations \eqref{DVqem}
are generally interpreted in a low energy effective action framework:
the gauge invariant composite fields $S_{I}$ are assumed to be good
low energy degrees of freedom, with low energy effective
superpotential given by $\wdv$. The conditions \eqref{DVqem} are then
the dynamical equations of motion in the low energy effective theory.
This interpretation is extremely difficult to justify from first
principles in the gauge theory, as stressed for example in
\cite{CDSW}. Our main result is to show that \eqref{DVqem} actually
follows from general consistency conditions in the gauge theory, in
some sense analogous to the Dirac quantization condition. More
precisely, we shall demonstrate that the vanishing of the differences
$(\partial_{s_{I}} - \partial_{s_{J}})\wdv$ follows from gauge
invariance, and that the vanishing of the sum
$\sum_{I}s_{I}\partial_{s_{I}}\wdv$ follows from a Ward identity. We
thus interpret the equations \eqref{DVqem} as being
\emph{non-dynamical identities}.

In the dual closed string theory formulation \cite{GV}, $\wdv$
\emph{is} a genuine effective superpotential associated with the
non-trivial background fluxes \cite{flux}. From this point of view,
deriving \eqref{DVqem} amounts to proving that the closed string
background proposed in \cite{GV}, and their generalizations based on
the geometric transition picture (see \cite{ferMM} for non-trivial
examples), are indeed the correct duals of the gauge theory. We thus
have a dual interpretation of the equations \eqref{DVqem}: they are
non-dynamical from the gauge theory, or open string point of view, and
dynamical from the closed string point of view. This is a nice new
example of a familiar phenomenon for dual formulations of the same
theory (the basic example being the electric-magnetic duality of the
Maxwell's equations, under which the Bianchi identities and the
dynamical equations of motion are exchanged).

A startling feature of the equations \eqref{DVqem} is that they
provide a unifying framework to derive essentially all the known exact
results in supersymmetric gauge theories. For example, they have been
used to obtain the Seiberg-Witten solutions of $\nn=2$ theories
\cite{CV}, to study Seiberg dualities \cite{Seidual}, or to find new
results on the quantum space of parameters in $\nn=1$ theories
\cite{ferQPS1,ferQPS2,phase1,phase2}. In some cases, the same results
can also be obtained by using various techniques (S-duality,
singularity theory, integrable systems, geometric engineering, mirror
symmetry, brane constructions, geometric transitions etc\ldots), but
at the expense of making several non-trivial assumptions. Actually,
there are very few derivations from first principles. The one
exception is Nekrasov's explicit summing-up of instanton series
\cite{nekrasov}, which yields the Seiberg-Witten prepotentials in
$\nn=2$ super Yang-Mills theories. The main application of the proof
of \eqref{DVqem} is to provide automatically assumption-free
derivations from first principles of all the above mentioned exact
results. For example, the derivation of the Seiberg-Witten $\nn=2$
prepotentials along these lines do not even assume that only
instantons contribute, as in Nekrasov's, put \emph{prove} it (see
Section 5). Another important application is to the mass gap and
confinement problems in $\nn=1$ super Yang-Mills. The main argument in
favor of confinement was given by Seiberg and Witten in their classic
papers \cite{SW}. The idea is to perturb a $\nn=2$ theory by adding a
tree level superpotential. The argument relies on two main hypotheses:
the solution of the $\nn=2$ theory is correct; the $\nn=1$ theory
obtained by perturbing $\nn=2$ creates a mass gap. We are able to
waive these hypotheses in Section 5.

The paper is organized as follows. In Section 2, we review some useful
concepts and results, including the work of Cachazo, Douglas, Seiberg
and Witten \cite{CDSW} on generalized Konishi anomaly equations. These
equations are used in Section 3 to study, for any adjoint field $X$ in
the theory, the quantum characteristic function
\be\label{qcf} \F(z) = \bigl\langle\det (z-X)\bigl\rangle\, .\ee
Our strategy is to show that gauge invariance puts general constraints
on $\F$ that turn out to be equivalent to
$(\partial_{S_{I}}-\partial_{S_{J}})\wdv = 0$. In Section 4, we find a
Ward identity that provides the missing equation in the form
$\sum_{I}S_{I}\partial_{S_{I}}\wdv = 0$. We also comment on the
possible existence of Kovner-Shifman vacua \cite{KSvac}. Finally in
Section 5 we apply the results to the mass gap and confinement
problems, giving en passant the derivation of the Seiberg-Witten
solution for $\nn=2$. Our arguments are completely general, but for
concreteness we mainly focus on the $\nn=1$ theory with gauge group
$\uN$, one adjoint matter superfield $X$ and $\Nf\leq 2N$ fundamental
flavors. We also give details on the case with two adjoint fields,
because it illustrates nicely some non-trivial aspects of the proof.

\section{Introductory material and discussion}
\setcounter{equation}{0}
\subsection{The quantum effective superpotential}

A most important object in four dimensional $\nn=1$ supersymmetric
gauge theories is the quantum effective superpotential $\wl$, which
determines the F-term part of the quantum effective action. It is
defined by a path integral over the vector and ghosts supermultiplets
$V$ and the matter chiral multiplets $X$, with boundary conditions
determined by the choice of a supersymmetric vacuum $|0\rangle$,
\be\label{wlowdef} e^{i\int\!\d^{4}x\left( 2N\re\int\!\d^{2}\theta\,
\wl^{|0\rangle} (T_{K}) + D\text{-terms}\right)} = \int_{|0\rangle} \!
[\d V\d X]\, e^{i\int\!\d^{4}x\,\mathscr{L}(V,X;T_{K})} \, . \ee
The normalization factor $N$ is chosen for convenience to coincide
with the number of colors of the gauge group $\uN$. The $T_{K}$
are external chiral superfields, with lowest components $t_{K}$, that
couple to gauge invariant operators ${\cal O}_{K}$ in such a way that
\be\label{wlder} \langle 0| {\cal O}_{K}|0\rangle = 
\frac{\partial\wl^{|0\rangle}}{\partial t_{K}}\,\cdotp\ee
In general, the knowledge of $\wl$ is equivalent to the knowledge of
the full set of chiral operator vacuum expectation values in the
theory.

For example, consider the $\uN$ gauge theory with one adjoint $X$,
$\Nf\leq 2N$ flavors of fundamentals $Q_{f}$ and anti-fundamentals
$\tilde Q^{f}$ chiral multiplets, and tree level lagrangian
\begin{multline}
\label{Lagdef}\mathscr L = \frac{1}{4\pi} \im\Bigl[ \tau_{\text{UV}}
\int\!\d^{2}\theta\, \Trf W^{\alpha}W_{\alpha}\Bigr] \\+
2N\re\int\!\d^{2}\theta\Bigl(\Trf\wt(X) + {}^{T}\tilde Q^{f}m_{f}^{\
f'}(X)Q_{f'}\Bigr)+\text{D-terms}\, . \end{multline}
The traces $\Trf$ are in the fundamental representation. The bare
gauge coupling $\tau_{\text{UV}}$ at scale $\La_{\text{UV}}$ is given
in terms of the dynamically generated scale $\La$ by
\be\label{tauren}2i\pi\tau_{\text{UV}} =
\ln\Bigl(\frac{\La}{\La_{\text{UV}}}\Bigr)^{2N-\Nf}\, .\ee
The one exception is when $\Nf=2N$, in which case $\tau_{\text{UV}}=
\tau$ is not renormalized. The derivative of the tree-level
superpotential is taken to be of the form
\be\label{wtp}\wt'(X) = \sum_{k = 0}^{d} g_{k} X^{k} = g_{d}
\prod_{I=1}^{d}(X-a_{I})\, .\ee
The polynomials
\be\label{mffpexp} m_{f}^{\ f'}(X) = \sum_{q\geq 0}m_{f,q}^{\
f'}X^{q}\ee
are such that
\be\label{detm}{\det}_{f,f'} m(X) = U_{0}\prod_{Q=1}^{\ell}(X-b_{Q})\,
.\ee
The most general vacuum 
\be\label{cavacua} |N_{1},k_{1};\ldots
;N_{d},k_{d};\nu_{1};\ldots;\nu_{\ell}\rangle\ee
is specified by the numbers $N_{I}$ and $\nu_{Q}$, $\nu_{Q}=0$ or 1,
of eigenvalues of the matrix $X$ that classically are equal to $a_{I}$
and $b_{Q}$, with pattern of gauge symmetry breaking $\uN\rightarrow
{\rm U}(N_{1})\times\cdots\times {\rm U}(N_{d})$, and by the integers
$k_{I}$, defined modulo $\mathbb Z_{N_{I}}$, that correspond to low
energy chiral symmetry breaking. In particular,
\be\label{sum}\sum_{I=1}^{d}N_{I}+\sum_{Q=1}^{\ell}\nu_{Q} = N\, .\ee
In any vacuum $|N_{I},k_{I};\nu_{Q}\rangle$, the relations
\eqref{wlder} read
\begin{align}\label{wlder2}
&\frac{1}{k+1}\bigl\langle \Trf X^{k+1}\bigr\rangle =
\frac{\partial\wl}{\partial g_{k}}\,\cvp\quad \bigl\langle
{}^{T}\tilde Q^{f}X^{q}Q_{f'}\bigr\rangle = \frac{\partial\wl}{\partial
m_{f,q}^{\ f'}}\,\cvp\\ \label{wlder2bis}
&\langle S\rangle = \frac{\partial\wl}{\partial\ln\La^{2N-\Nf}}\,\cvp
\end{align}
where we have introduced the glueball superfield
\be\label{gluedef} S=-\frac{\Trf W^{\alpha}W_{\alpha}}{16\pi^{2}N}\, 
\cdotp\ee
When $\Nf=2N$, $\ln\La^{2N-\Nf}$ is replaced by $2i\pi\tau$ in 
\eqref{wlder2bis}.

By a standard non-renormalization theorem, $\wl$ is not corrected in
perturbation theory, but there are in many cases non-trivial
non-perturbative corrections, typically given by infinite sums over
(fractional) instantons. The calculation of $\wl$ is thus a difficult
strong coupling problem.

\subsection{The Dijkgraaf-Vafa approach}

Dijkgraaf and Vafa made the remarkable conjecture \cite{DV} that the
calculation of $\wl$ in a large class of $\nn =1$ supersymmetric gauge
theories could be reduced to a simple recipe. The recipe goes in two
separate steps.
\subsubsection{Step one: the matrix model}
First one introduces a zero-dimensional gauged matrix model whose
potential coincides with the tree-level superpotential of the gauge
theory. For example, for the model \eqref{Lagdef}, we use the
one-matrix model coupled to flavors, for which the partition function
$\mathcal F$ is an integral over a $n\times n$ hermitian matrix $X$
and $n$-vectors $\tilde Q^{f}$ and $Q_{f'}$,
\be\label{MMdef} e^{n^{2}\Fb/s^{2}}= \mathscr N\int\!\d X \d Q\d \tilde
Q\, e^{-\frac{n}{s}\Trf\wt(X)-\frac{n}{s}{}^{T}\tilde Q^{f}m_{f}^{\
f'}(X)Q_{f'}}\, .\ee
The normalization factor $\mathscr N$ is independent on the couplings
\be\label{tKdef} t_{K}=(g_{k},m_{f,q}^{\ f'})\ee
in the tree-level action and will be discussed later. The size $n$ of
the matrix and the vectors is a dummy variable and is not related to
the number of colors $N$ in the gauge theory. We are actually
interested in the $n\rightarrow\infty$ limit, or more precisely in the
sum over sphere and disk diagrams around a particular classical
($s\rightarrow 0$) solution, which yields an expansion of the form
\be\label{Flargen} \mathcal F = \Fp + \frac{s}{n}\Fd + \mathcal
O(1/n^{2})\, .\ee
The classical solutions are classified by the integers $n_{I}$ and
$\nu_{Q}$, $\nu_{Q}=0$ or 1, where $n_{I}$ and $\nu_{Q}$ give the
number of eigenvalues of $X$ equal to $a_{I}$ and $b_{Q}$ in the
$s\rightarrow 0$ limit. In the $n\rightarrow\infty$ limit, the
partition function, obtained by expanding around a particular matrix
model vacuum 
\be\label{mmvacua} |n_{1},\ldots
,n_{d};\nu_{1},\ldots,\nu_{\ell}\rangle\, ,\ee
is a natural function of the filling fractions
\be\label{ff} s_{I}=s\,\frac{n_{I}}{n}\ee
and of the $\nu_{Q}$. Actually, since the sphere partition function
$\Fp$ is insensitive to the coupling to the fundamental flavors, it
depends on the $s_{I}$ only, whereas the disk partition function $\Fd$
depends on both the $s_{I}$ and the $\nu_{Q}$.

We then introduce the Dijkgraaf-Vafa superpotential \cite{DV,addfla}
\be\label{DVpot} \wdv(s_{I};t_{K}) =
-\sum_{I}N_{I}\frac{\partial\Fp}{\partial s_{I}}-\Fd\, .\ee
The ``weak'' Dijkgraaf-Vafa conjecture states that
\be\label{weakDV} \frac{\partial\wdv}{\partial g_{k}}=
\frac{1}{k+1}\bigl\langle \Trf X^{k+1}\bigr\rangle\, ,\quad
\frac{\partial\wdv}{\partial m_{f,k}^{\ f'}} = \bigl\langle 
{}^{T}\tilde Q^{f}X^{k}Q_{f'}\bigr\rangle\, ,\ee
where the $\nu_{Q}$ parametrizing the matrix model \eqref{mmvacua} and
gauge theory \eqref{cavacua} vacua are taken to be the same. The
equations \eqref{weakDV} determine an infinite set of chiral operators
vacuum expectation values in terms of a finite number of undetermined
constants $s_{I}$. These matrix model filling fractions are identified
in the gauge theory with the gluino condensates in the
$\text{U}(N_{I})$ factors of the low energy gauge group,
\be\label{sI} s_{I}=\langle S_{I}\rangle\, ,\ee
or are taken to be zero if $N_{I}=0$ (as explained in \cite{CDSW}, the
operators $S_{I}$ are defined as in \eqref{gluedef} with the insertion
of a suitable projection operator).
\subsubsection{Step two: the extremization of $\wdv$}
The ``strong'' Dijkgraaf-Vafa conjecture states that 
\be\label{wlowDV} \wl = \wdv (s_{I}^{0})\, ,\ee
where the $s_{I}^{0}$ are the particular values of the $s_{I}$ that
solve the quantum equations of motion \eqref{DVqem}.

The function $\wdv$ splits into a part $\wvy$ that contains all the
$t_{K}$-independent terms in $\wdv$ and the $t_{K}$-dependent part
that we denote $\wmm$,
\be\label{wDVsplit}\wdv(s_{I};t_{K}) = \wvy(s_{I}) + \wmm(s_{I};t_{K})\,
,\ee
It is important to realize that the weak Dijkgraaf-Vafa conjecture
does not depend on the form of $\wvy$, since any $t_{K}$-independent
term drops from \eqref{weakDV}. On the other hand, $\wvy$ plays a
crucial r\^ole in the equations \eqref{DVqem}. A central result of the
present work will be to derive, in Sections 3 and 4, the form of
$\wvy$ from first principles.

With our definition \eqref{MMdef}, $\wvy$ depends on the normalization
factor $\mathscr N$. A priori, this normalization factor may be a
complicated function of $s$ and of the dynamically generated scale
$\La$. For example, for the model \eqref{Lagdef}, we may postulate
$\mathscr N\propto \La^{a} s^{b}$ with $a+3b = -n^{2}-2n\Nf$ for
dimensional reasons. The correct choice (that we will be able to
justify later) turns out to be
\be\label{Nc1}\mathscr N \propto \La^{-n^{2}+n\Nf}s^{-n\Nf}\, .\ee
There can be an additional purely numerical factor (independent of the
$t_{K}$, $s_{I}$ or $\La$), that can be used to set a convenient
normalization of $\La$. 

A last important comment is that with the choice \eqref{Nc1},
\eqref{MMdef} and \eqref{DVpot} would imply that $\wvy =
-\sum_{I}N_{I}s_{I}\ln s_{I}$, which is slightly misleading. The
correct formula is
\be\label{VYspe} \wvy(s_{I}) =
-\sum_{I}s_{I}\ln s_{I}^{N_{I}}\, .\ee
With this prescription, solving \eqref{DVqem} automatically yields the
full set of chirally asymmetric vacua \eqref{cavacua} labeled by the
integers $k_{I}$. Let us note that $\wvy$ is the sum of the
independent pure gauge theory contributions postulated in \cite{VY}
for each factor $\text U(N_{I})$ of the low energy gauge group. For
this reason, it is often called the Veneziano-Yankielowicz part of the
Dijkgraaf-Vafa superpotential.

\subsection{The generalized Konishi anomaly equations}

The weak form of the Dijkgraaf-Vafa conjecture was explained in a
seminal paper by Cachazo, Douglas, Seiberg and Witten
\cite{CDSW,seifla}.\footnote{Unfortunately, a proper non-perturbative
discussion of the arguments in \cite{CDSW} has not yet appeared in the
literature. In the present paper we focus on the strong form of the
Dijkgraaf-Vafa conjecture, assuming that the weak form of the
conjecture, or equivalently the results of \cite{CDSW}, are correct.}
These authors showed that the loop equations of the matrix model, that
determine all the matrix model correlators as a function of the
filling fractions $s_{I}$, were in one-to-one correspondence with a
set of generalized Konishi anomaly equations in the gauge theory, the
precise form of the correspondence being equivalent to \eqref{weakDV}.

\FIGURE{\epsfig{file=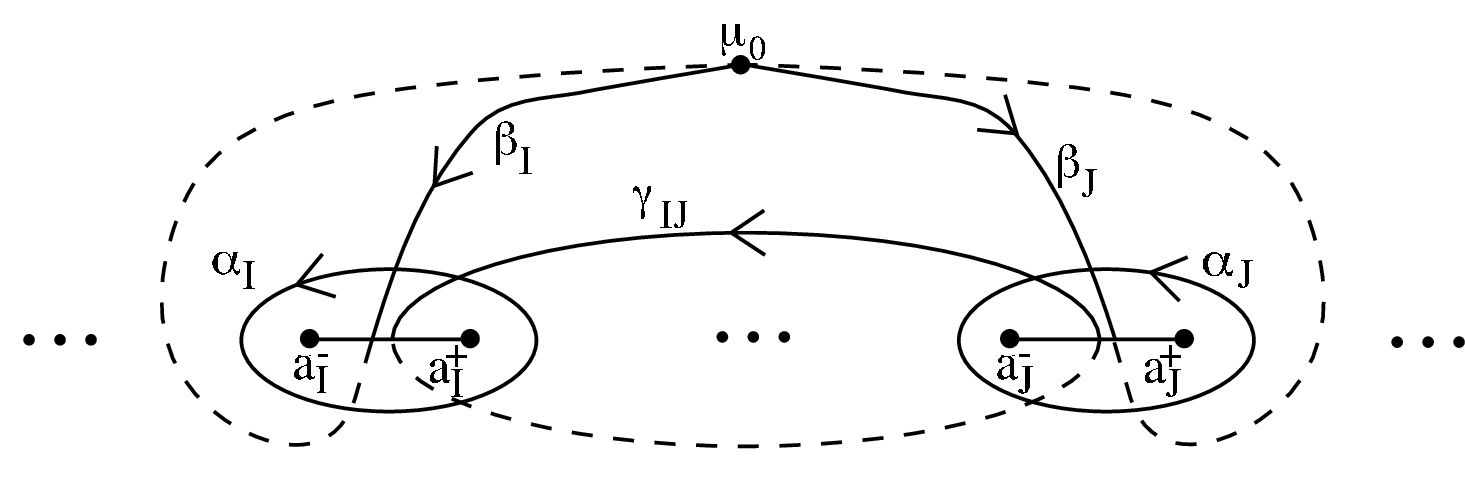,width=14cm}
\caption{The non-compact two-sheeted Riemann surface $\mathcal C$,
with the contours $\alpha_{I}$, $\beta_{I}$ and $\gamma_{IJ}$ used in
the main text. The contours $\beta_{I}$ are defined modulo an integer
multiple of $\alpha_{I}$. The points $z=\mu_{0}$ on the first and
second sheets are taken to infinity.
\label{fig1}}}

The consequences of the generalized Konishi anomaly equations, and in
particular of the equations \eqref{weakDV}, for the model
\eqref{Lagdef} can be summarized as follows. Let us introduce the
generating functions for the $\langle\Trf X^{k}\rangle$, $\langle\Trf
W^{\alpha}W_{\alpha}X^{k}\rangle$ and $\langle {}^{T}\tilde
Q^{f'}X^{q}Q_{f}\rangle$,
\begin{align}\label{GF}\mathscr R(z) &=
\Bigl\langle\Trf\frac{1}{z-X}\Bigr\rangle\, \cvp\\
\label{GF1} \mathscr S(z)&= -\frac{1}{16\pi^{2}
N}\Bigl\langle\Trf\frac{W^{\alpha}W_{\alpha}}{z-X}\Bigr\rangle\,
\cvp\\ \label{GF2}\mathscr G_{f}^{\ f'}(z) &=
\Bigl\langle{}^{T}\tilde Q^{f'}\frac{1}{z-X}Q_{f}\Bigr\rangle\,\cvp
\end{align}
and the convenient combination
\be\label{Ydef}\Y (z)= \wt'(z) - 2\S(z)\, .\ee
The function $\mathscr Y$ satisfies a quadratic equation
\be\label{Kon1} \Y(z)^{2}=\wt'(z)^{2} - \Delta(z)\, ,\ee
where $\Delta$ is a polynomial of degree $\deg\Delta\leq d-1=\deg\wt'
-1$. In a vacuum with low energy gauge group $\text
U(N_{1})\times\cdots\times\text U(N_{\tilde d})$, $1\leq\tilde d\leq
d$, the following factorization in terms of a degree $d-\tilde d$
polynomial $H_{d-\tilde d}$ takes place,
\be\label{fact1} \wt'^{2}(z) - \Delta (z) = H_{d-\tilde d}^{2}(z)
\prod_{I=1}^{\tilde d}(z-a_{I}^{-})(z-a_{I}^{+})\, ,\ee
and thus $\mathscr Y$ and $\S$ are meromorphic functions on the genus
$\tilde d-1$ hyperelliptic surface
\be\label{RS} \mathcal C:\ y^{2} = 
\prod_{I=1}^{\tilde d}(z-a_{I}^{-})(z-a_{I}^{+})\, .\ee
This two-sheeted Riemann surface, with some contours defined on it, is
depicted in Figure \ref{fig1}. The polynomial $\Delta$, and thus the
surface $\mathcal C$, is completely determined by the factorization
condition \eqref{fact1} up to $\tilde d$ parameters that can be
conveniently chosen to be the $\alpha_{I}$-period integrals of $\S\d
z$,
\be\label{idS}\frac{1}{2i\pi}\oint_{\alpha_{I}}\!\S\d z = s_{I}\,
.\ee
In particular, we have
\be\label{sumsI}\sum_{I}s_{I} = s = \langle S\rangle\, ,\ee
the expectation value of the glueball superfield defined in
\eqref{gluedef}. In the classical limit, $s_{I}\rightarrow 0$ for all
$I$ and we have $a_{I}^{-}\simeq a_{I}^{+}\simeq a_{I}$. Equations
\eqref{Ydef} and \eqref{Kon1} also show that $\S(z)$ is the sphere
matrix model resolvent,
\be\label{Seq} \S(z) =
\Bigllangle\frac{s}{n}\Trf\frac{1}{z-X}\Bigrrangle_{\text{sphere}}\,
\cvp\ee
where the double bracket notation is used for the matrix model
expectation values (which are clearly different in general from the
expectation values in the gauge theory). The parameters $s_{I}$
introduced in \eqref{idS} are thus the same as the filling fractions
\eqref{ff} of the matrix model.

The generating functions $\mathscr G_{f}^{\ f'}$ and $\R$ are also
meromorphic functions on the curve \eqref{RS}. We shall use the gauge
theory resolvent $\R$ extensively in the following, so let us describe
its structure in details. Let us introduce
\be\label{defpsi} \psi_{I}(z) = \frac{\partial\S(z)}{\partial
s_{I}}\,\cdotp\ee
It is straightforward to show that the $\psi_{I}\d z$ form a basis of
log-normalizable holomorphic one-forms on $\mathcal C$,
\be\label{psiIform} \psi_{I} = \frac{p_{I}}{y}\,\cdotp\ee
The polynomials $p_{I} = z^{\tilde d - 1}+\cdots$ are fixed by the
constraints
\be\label{psinorm} \frac{1}{2i\pi}\oint_{\alpha_{J}}\psi_{I}\,\d z =
\delta_{IJ}\, .\ee
The gauge theory resolvent is then given by
\be\label{Req} \R(z)=\sum_{I}N_{I}\psi_{I}(z) +g_{\text d}(z)\,
,\ee
where $g_{\text d}$ is the disk matrix model resolvent (obtained by
summing disk diagrams only),
\be\label{diskres} g_{\text d}(z) =
\Bigllangle\Trf\frac{1}{z-X}\Bigrrangle_{\text{disk}}\, .\ee
If we define the ``classical'' sheet of the Riemann surface \eqref{RS}
by the condition
\be\label{clasymp}\R(z)\underset{z\rightarrow\infty}{\sim}\frac{N}{z}
\,\cvp\ee
then $\R$ has poles on the classical sheet at $z=b_{Q}$ with residue
$\nu_{Q}$ if $\nu_{Q}=1$, and poles on the second sheet at $z=b_{Q}$
with residue $1-\nu_{Q}$ if $\nu_{Q}=0$. We also have the useful
identities
\be\label{idR} \frac{1}{2i\pi}\oint_{\alpha_{I}}\!\R(z)\,\d z =
N_{I}\, .\ee
For completeness, let us finally mention that
\be\label{Gff} \mathscr G_{f}^{\ f'}(z) = \Bigllangle{}^{T}\tilde
Q^{f'}\frac{1}{z-X}Q_{f}\Bigrrangle_{\text{disk}}\, . \ee
The above equations determine all the $\langle\Trf X^{k}\rangle$,
$\langle W^{\alpha}W_{\alpha}\Trf X^{k}\rangle$ and $\langle
{}^{T}\tilde Q^{f'}X^{q}Q_{f}\rangle$ in terms of the $s_{I}$. Note
that taking into account \eqref{defpsi}, \eqref{Req} and \eqref{Gff}
are perfectly consistent with \eqref{weakDV} and \eqref{DVpot}.

\subsection{Another formulation of the quantum equations of motion}

A last important technical result that we need is an elegant
reformulation of the quantum equations of motion \eqref{DVqem} as
constraints on the period integrals of $\R\d z$ \cite{phase2}. Using the
open contours $\beta_{I}$ defined in Figure 1, and with the
understanding that $\mu_{0}$ is taken to infinity, the equations
\eqref{DVqem} are equivalent to
\be\label{qem2}\frac{1}{2i\pi}\int_{\beta_{I}}\!\R(z)\,\d z =
\frac{1}{2i\pi}\ln\frac{U_{0}\La^{2N-\Nf}}{\mu_{0}^{2N-\ell}}
+ k_{I}\, ,\ee
where the $k_{I}$s are integers defined modulo $N_{I}$. In particular,
the periods on the compact cycles $\gamma_{IJ}=\beta_{I}-\beta_{J}$
are integers,
\be\label{intpe}\frac{1}{2i\pi}\oint_{\gamma_{IJ}}\!\!
\R(z)\,\d z=k_{I}-k_{J}\in\mathbb Z\, .\ee

Since this result is crucial for our purposes, we now present a short 
proof. The starting point is the so-called special geometry relations
of the matrix model,
\be\label{secder} \frac{\partial^{2}\Fp}{\partial s_{I}\partial s_{J}}
= \int_{\beta_{I}}\!\psi_{J}\,\d x + 2 \ln\frac{\mu_{0}}{\La}\,
\cvp\ee
as well as similar equations for the derivative of the disk partition 
function,
\be\label{gdperiod}\frac{\partial\Fd}{\partial s_{I}} =
\int_{\beta_{I}}\! g_{\mathrm d}\,\d x +
\ln\bigl(\mu_{0}^{2\sum_{Q=1}^{\ell}\nu_{Q} - \ell}/U_{0}\bigr) \,
.\ee
The details on the derivation of these matrix model relations can be
found for example in \cite{ferrev}. Equations \eqref{DVpot} and
\eqref{Req} then yield
\be\label{DVpotder}\frac{\partial\wdv}{\partial s_{I}} =
-\int_{\beta_{I}}\!\R(z)\,\d z -
\ln\frac{\mu_{0}^{2N-\ell}}{U_{0}\La^{2N-\Nf}}\, \cdotp\ee
This would correspond to a Veneziano-Yankielowicz term $-N_{I}\ln
s_{I}$ in $\partial\wdv/\partial s_{I}$, whereas the correct term is
$-\ln s_{I}^{N_{I}}$ as in \eqref{VYspe}. Taking into account this
subtlety, we obtain \eqref{qem2}. The $k_{I}$ are defined modulo
$N_{I}$ because the contours $\beta_{I}$ are defined modulo an integer
multiple of $\alpha_{I}$ (the transformation
$\beta_{I}\rightarrow\beta_{I}+\alpha_{I}$ is induced by continuously
permuting $a_{I}^{-}$ and $a_{I}^{+}$, which obviously does not change
the curve $\mathcal C$ nor any gauge theory correlator, and thus
yields the same solution). 

The most natural form of \eqref{qem2} is
\be\label{qem3}\mu_{0}^{2N-\ell}\exp\int_{\beta_{I}}\!\R (z)\,\d z =
U_{0}\La^{2N-\Nf}\, .\ee
From the above discussion, we know that there are automatically
$\prod_{I=1}^{\tilde d}N_{I}$ distinct solutions to these equations,
labelled by the $\mathbb Z_{N_{I}}$-valued integers $k_{I}$.

\subsection{Discussion}

The Konishi anomaly equations can be generalized to a very large class
of models and put very powerful constraints on the set of chiral
operators expectation values. However, any choice of the parameters
$s_{I}$ yields a perfectly sensible solution of the equations. For
instance, the choice $s_{I}=0$ for all $I$ corresponds to no quantum
corrections at all to $\wl$. Clearly, the anomaly equations do not
address directly the problem of the strongly coupled gauge dynamics.
This was explained in details already in the original paper
\cite{CDSW}. Nevertheless, these equations \emph{do} provide some
important indirect information, since we know that the true answer
must correspond to \emph{some} particular values of the $s_{I}$. In
particular, we know that $\S$ and $\R$ are functions defined on the
two-sheeted surface \eqref{RS}, since this is true for \emph{all}
values of the $s_{I}$. The knowledge of this analytic structure will
be of great help in the following.

Unlike the Konishi anomaly equations, the strong part of the
Dijkgraaf-Vafa conjecture deals directly with genuine non-perturbative
effects in the gauge theory, including infinite series of (fractional)
instanton contributions and chiral symmetry breaking. It is also
closely related to the problems of the mass gap and confinement in
$\nn=1$ theories, as we shall review in Section 5.

The original formulation of the conjecture in terms of the
extremization of the function $\wdv$ suggests that $\wdv$ is a low
energy effective superpotential and the equations \eqref{DVqem} purely
dynamical. It would clearly be very difficult to devise a proof along
these lines. For example, there is no convincing argument showing that
the glueball fields $S_{I}$ are the good low energy variables.
Actually, this interpretation is logically disconnected from the
Dijkgraaf-Vafa conjecture, and it is even unclear whether it is fully
correct or not.

The equations \eqref{qem2} or \eqref{qem3}, even though they are
mathematically strictly equivalent to the original Dijkgraaf-Vafa
formulation in terms of a superpotential, are beautifully simple and
offer an attractive alternative starting point. At the very least,
they suggest an entirely different interpretation. In particular, the
equations \eqref{intpe} look like simple quantization conditions,
which are often derived from general consistency constraints. We show
in the next Section that this is indeed the case, \eqref{intpe} being
a consequence of gauge invariance. The full set of equations
\eqref{qem3} then follows from an additional constraint, which is a
simple Ward identity that we derive in Section 4.

The relation between \eqref{intpe} and gauge invariance is not totally
unexpected. \emph{Assuming} that confinement is realized in the $\text
U(N_{I})$ factors of the low energy gauge group through the
condensation of dyons of unit magnetic charge (the usual 't~Hooft's
mechanism), one can show that the integers $k_{I}$ in \eqref{qem2}
coincide with the electric charges of the dyons \cite{phase1}. The
Dirac-Schwinger-Zwanziger quantization condition, which is well-known
to be a direct consequence of gauge invariance, is then equivalent to
\eqref{intpe}. Our main task in Section 3 is to present a simple 
derivation of the same result that is free from any assumption on the 
low energy behaviour of the theory.

\subsection{Other models}

An important property of \eqref{intpe} is its universality. Even
though the details of the Dijkgraaf-Vafa prescription (matrix model,
definition of $\wdv$) are model-dependent, \emph{the extremization of
the Dijkgraaf-Vafa superpotential always yields quantization conditions
of the form \eqref{intpe}, with $\R$ being defined as in \eqref{GF} to
be the resolvent associated with any adjoint chiral superfield in the
theory.} This fact will find a simple explanation in the next Section,
but we wish to illustrate it using elementary methods on a
representative example.

Let us introduce the theory with two adjoints $X$ and $Y$ and a
tree-level superpotential of the form
\be\label{wtree2ad}\wt(X,Y) = V_{1}(X) + V_{2}(Y) - \mu XY\, ,\quad
V_{1}' = P\, ,\ V_{2}'=Q\, .\ee
This case is particularly interesting, because there are now two natural
gauge theory resolvents,
\be\label{RXRY}\R^{X}(z) = \Bigl\langle\Trf\frac{1}{z-X}\Bigr\rangle\,
\cvp\quad \R^{Y}(z) = \Bigl\langle\Trf\frac{1}{z-Y}\Bigr\rangle\,
\cvp\ee
for which constraints of the form \eqref{qem3} can be expected to be
valid. This is non-trivial, because the constraints are then
redundant, and not obviously consistent with each other.

The matrix model we have to consider is
\be\label{twoMM} e^{n^{2}\Fb/s^{2}}= \mathscr N\int\!\d X \d Y
\, e^{-\frac{n}{s}\Trf\wt(X,Y)}\, ,\ee
with a Dijkgraaf-Vafa superpotential 
\be\label{wdv2mm} \wdv = -\sum_{I}N_{I}\frac{\partial\Fp}{\partial
s_{I}}\,\cdotp\ee
It is then possible to show that, provided the normalization constant
is chosen to be
\be\label{norm2MM}\mathscr N \propto \La^{-n^{2}/2}s^{-n^{2}/2}\, ,\ee
the equations of motion \eqref{DVqem} take a form similar to
\eqref{qem3}. In particular, the quantization conditions
\be\label{qc2MMa} \frac{1}{2i\pi}
\oint_{\gamma_{IJ}^{X}}\!\!\R^{X} (z)\,\d z\in\mathbb
Z\ee
and
\be\label{qc2MMb}
\frac{1}{2i\pi}\oint_{\gamma_{IJ}^{Y}}\!\!\R^{Y} (z)\,\d z\in\mathbb Z\ee
turn out to be equivalent to each other. The choice \eqref{norm2MM}
for $\mathscr N$ is also necessary for the standard interpretation in
terms of Veneziano-Yankielowick terms to be valid. The contours
$\gamma_{IJ}^{X}$ and $\gamma_{IJ}^{Y}$ are similar to the contours
$\gamma_{IJ}$ of Figure 1, but with the $\gamma_{IJ}^{X}$
(respectively $\gamma_{IJ}^{Y}$) going through the cuts of the
resolvent $\R^{X}$ ($\R^{Y}$). The proof of \eqref{qc2MMa} and
\eqref{qc2MMb} uses the generalized Konishi anomaly equations (they
have not appeared in the literature for the model \eqref{wtree2ad},
but they are straightforward to derive) and the special geometry
relations for the matrix model that are derived in \cite{ferMM}. Full
details are given in \cite{ferrev}.

\section{The quantization conditions}
\setcounter{equation}{0}
\subsection{Presentation of the problem}

Let $M$ be an arbitrary adjoint chiral superfield. For example, in the
model \eqref{Lagdef}, we may consider any power of the elementary
field $X$; in the model \eqref{wtree2ad}, any combination of the form
$\prod_{k} X^{p_{k}}Y^{p_{k}}$ would qualify. The resolvent for $M$ is
defined by
\be\label{resdef}\R(z) =
\Bigl\langle\Trf\frac{1}{z-M}\Bigr\rangle\,\cdotp\ee
The classical resolvent has the form
\be\label{rescl}\R_{\text{cl}}(z) =
\sum_{I=1}^{D}\frac{N_{I}}{z-a_{I}}\,\cvp\ee
where the integers $N_{I}$, $\sum_{I}N_{I}=N$, give the number of
eigenvalues of $M$ that are equal classically to $a_{I}$. Quantum
mechanically, we assume that $\R$ is a multi-sheeted analytic
function. On the classical sheet, which is defined by the property
that the large $z$ asymptotics is the same as in the classical theory,
\be\label{psasymp}\R(z)\underset{z\rightarrow\infty}{\sim}\frac{N}{z}
\,\cvp\ee
some of the poles (say the $a_{I}$ for $1\leq I\leq d$) in
\eqref{rescl} are replaced by square root branch cuts, and the others
(the $a_{Q}$ for $Q\geq d+1$) remain poles that may be shifted from
their classical value $a_{Q}$ to a quantum value $b_{Q}$. The analytic
structure is thus similar to the one given in Figure \ref{fig1} (in
general there could be additional branch cuts on the second sheet, and
the classical sheet could be glued to many different sheets by the
square root branch cuts; we shall work with an analytic structure of
the form given by Figure \ref{fig1}, because the other cases amount to
straightforward generalizations). As explained in Section 2, the above
assumptions on $\R$ can be derived from the generalized Konishi
anomalies in many cases (including the cases that are relevant for our
purposes), and so they are not restrictive.

The classical limit \eqref{rescl} and the analytic structure of $\R$
imply trivially that
\be\label{genNI}\frac{1}{2i\pi}\oint_{\alpha_{I}}\!\R(z)\,\d z
=N_{I}\in\mathbb Z\, .\ee
We want to prove that gauge invariance implies that the other period
integrals of $\R\d z$ must be quantized as well,
\be\label{genqc}\frac{1}{2i\pi}\oint_{\gamma_{IJ}}\!\!\R(z)\,\d z
\in\mathbb Z\, .\ee
\subsection{On gauge invariance}\label{Sgi}

In \cite{CDSW}, the single-trace gauge invariant chiral operators in
the theory with no flavors were classified (a similar classification
can be achieved in many other models). The full set is given by $\Trf
X^{k}$, $\Trf X^{k}W^{\alpha}$ and $\Trf X^{k}W^{\alpha}W_{\alpha}$.
We call the ring generated by these elements the reduced chiral ring
of the theory. Because of the factorization of chiral operators
correlation functions, the structure of this ring is fully determined
by the knowledge of the functions $\R(z)$ and $\S(z)$ defined in
\eqref{GF} and \eqref{GF1}. 

The most general element of the reduced chiral ring can be written as
a \emph{finite} sum of \emph{finite} products of single trace chiral
operators. This does not exhaust the full set of interesting chiral
operators. In particular, $\R(z)$ and $\S(z)$ themselves are not in
this ring. More generally, it is natural to consider operators of the
form
\be\label{opdef} O(z) = \sum_{k\geq K} O_{k}z^{-k}\,
,\ee
where the $O_{k}$ are elements of the reduced chiral ring. The series 
defining the expectation values
\be\label{opvev}\mathscr O(z) = \sum_{k\geq K}\langle O_{k}\rangle
z^{-k}\ee
are assumed to have a non-zero radius of convergence around
$z=\infty$. Operators like \eqref{opdef} are usually introduced
because they are generating functions for the $\langle O_{k}\rangle$.
This interpretation comes from the expansion at large $z$. In
particular, when $z$ is in the disk on convergence, there is little to
learn in considering $O(z)$ instead of the individuals $O_{k}$.
However, we shall see that considering a priori $O(z)$ for any value
of $z$ introduces some interesting subtleties.

When $z$ is not in the disk of convergence, $\mathscr O(z)$ is defined
by analytic continuation. The result is in general a multivalued
function on the $z$-plane. This multivaluedness is not obviously
consistent with gauge invariance. Gauge invariance is manifest in an
expansion like \eqref{opdef}, but this expansion is only valid on a
single sheet, and in general a similar expansion does not exist on the
other sheets. To understand how the problem can arise, let us focus on
the case where $\mathscr O(z)$ is built entirely in terms of a single
adjoint field $M$ in the theory. The operators $O_{k}$ are then
polynomials in the $\Trf M^{q}$, and thus $\mathscr O(z)$ depends only
on the eigenvalues $m_{i}$ of $M$ on the sheet where \eqref{opvev} is
valid, and thus on all the other sheets by analytic continuation,
\be\label{eigendep}\mathscr O(z) =\mathscr O(z;m_{1},\ldots,m_{N})\,
.\ee
The fundamental constraint from gauge invariance is then that
\emph{$\mathscr O(z;m_{i})$ must be a symmetric function of the
$m_{i}$.} If $\sigma\cdot$ is the operator implementing the gauge
transformation associated with the permutation $\sigma\in
S_{N}\subset\uN$ of the eigenvalues, we must have
\be\label{permute} (\sigma\cdot\mathscr O)(z;m_{i})= \mathscr
O(z;m_{\sigma(i)}) = \mathscr O(z;m_{i})\, .\ee
This constraint is obviously satisfied when the series \eqref{opvev}
converges, but is \emph{not} insured for the analytic continuations.

To illustrate this point, let us consider an extremely simple
classical toy example,
\be\label{toy} O(z) = \Bigl(\Trf\sqrt{z-M}\Bigr)^{2}= N^{2}z - N\Trf M
+\frac{1}{4z}\bigl((\Trf M)^{2}-N\Trf M^{2}\bigr) + \cdots\, .\ee
In terms of the eigenvalues, we have
\be\label{toyeigen}\mathscr O(z;m_{i}) = 
\sum_{i, j}\sqrt{(z-m_{i})(z-m_{j})}\, .\ee
The function $\mathscr O(z;m_{i})$ has square root branch cuts that
join pairs of eigenvalues. Let us start from a value of $z$ for which
the series in the right hand side of \eqref{toy} converges, and let us
consider the analytic continuation through the cut between $m_{i_{0}}$
and $m_{j_{0}}$. This analytic continuation amounts to changing the
sign of the term $\sqrt{(z-m_{i_{0}})(z-m_{j_{0}})}$ in
\eqref{toyeigen}, and is thus clearly inconsistent with the symmetry
of $\mathscr O$ under the permutation of $m_{i_{0}}$ (or $m_{j_{0}}$)
with any other eigenvalue $m_{i}$ when $i\not\in\{i_{0},j_{0}\}$. This
shows that $O(z)$ defined by \eqref{toy} is not a good gauge invariant
operator for all values of $z$. We could of course compute it at large
$z$ in the sheet where it is gauge invariant, and use it as a
convenient book-keeping device for the $\Trf X^{k}$. However,
considering $\bigl(\Trf\sqrt{z-M}\bigr)^{2}$ for all values of $z$ is
inconsistent with gauge invariance.

In the general case, let us label by an index $a\geq 0$ the different
sheets of the analytic function $\mathscr O$, and write $\mathscr
O_{a}(z;m_{i})$ its value on the $a^{\text{th}}$ sheet. Suppose that
the $0^{\text{th}}$ sheet contains the disk of convergence of
\eqref{opvev}, and thus in particular that
\be\label{permutea} (\sigma\cdot\mathscr O_{0})(z;m_{i}) = \mathscr
O_{0}(z;m_{i})\, .\ee
If the action of the operator $\sigma\cdot$ is smooth, the same
equation must automatically be valid on any other sheet, by analytic
continuation. This shows that if there exists a sheet, say number $a$,
for which the permutation symmetry is violated,
\be\label{permuteb} (\sigma\cdot\mathscr O_{a})(z;m_{i}) \not = \mathscr
O_{a}(z;m_{i})\, ,\ee
then the gauge transformation $\sigma\cdot$ must be singular when
acting on $\mathscr O(z)$ for some values of $z$. For example, if the
$0^{\text{th}}$ and $a^{\text{th}}$ sheets are joined by a branch cut,
then $\sigma\cdot$ is clearly not well defined (two-valued) on this
cut. This inconsistency is similar to the problems that appear in the
presence of a Dirac string when the Dirac quantization condition is
not satisfied.

\subsection{The quantum characteristic function}
\subsubsection{Generalities}

Let us introduce the characteristic function of $M$,
\be\label{cfdef} \F(z) = \bigl\langle\det(z-M)\bigr\rangle\, ,\ee
with the basic property that
\be\label{RFrel} \frac{\F'(z)}{\F(z)} = \R(z)\, .\ee

By using
\be\label{iddet} \det (z-M) = z^{N}e^{\Trf\ln (1-M/z)}\, ,\ee
we can obtain an expansion of the form \eqref{opdef},
\be\label{detexp} \det(z-M) = z^{N} - \sum_{k\geq 1} F_{k} z^{N-k}\,
,\ee
where the elements of the reduced chiral ring $F_{k}=\frac{1}{k}\Trf
M^{k}+\cdots$ are homogeneous polynomials in the $\Trf M^{q}$ of
degree $k$ ($M$ being of degree one) whose form does not depend
explicitly on the number of colors $N$. For example,
\be\label{exFk}
\begin{split}
F_{1} &= \Trf M\, ,\\ F_{2} &= \frac{1}{2}\Trf M^{2} -
\frac{1}{2}\bigl(\Trf M\bigr)^{2}\, ,\\ F_{3} &= \frac{1}{3}\Trf M^{3}
- \frac{1}{2} \bigl(\Trf M\bigr) \bigl(\Trf M^{2}\bigr) +
\frac{1}{6}\bigl(\Trf M\bigr)^{3}\, ,\ \text{etc\ldots} \end{split}\ee
Picking a particular value of $N$ amounts to imposing relations in the
chiral ring,
\be\label{relcr} F_{N + q} = 0\, ,\quad q\geq 1\, ,\ee
that are equivalent to the fact that classically $\det(z-M)$ is a
polynomial of degree $N$. Using \eqref{relcr} recursively, one obtains
the standard indentities
\be\label{relcrS} \Trf M^{q} = P_{q,\text{cl}}\bigl( \Trf
M,\ldots,\Trf M^{N}\bigr)\, ,\quad q\geq 1\, ,\ee
where the $P_{q,\text{cl}}$ are homogeneous polynomials of degree $q$
depending only on the $N$ independent Casimir $\Trf M^{k}$, $1\leq
k\leq N$, of $\uN$.

The chiral ring relations \eqref{relcr} or \eqref{relcrS} can get
non-perturbative corrections,
\be\label{qrel} \Trf M^{q} = P_{q}\bigl( \Trf M,\ldots,\Trf
M^{N};\La\bigr)\, . \ee
The generalized Konishi anomaly equations put strong constraints on
the form of the polynomials $P_{q}$. In particular, if the filling
fractions $s_{I}$ are not all equal to zero, the quantum corrections
must be non-trivial. The equations \eqref{qrel} then imply that the
quantum characteristic function \eqref{cfdef} or \eqref{detexp} is not
a polynomial, but an infinite series, the non-polynomial terms
precisely generating the quantum corrections to the chiral ring.

\subsubsection{The integral representation}

The characteristic function $\F(z)$ is thus multivalued (this follows
from \eqref{RFrel} and the multivaluedness of $\R(z)$). The fact that
a correlator can be multivalued in a field theory is of course not an
inconsistency, and actually is a very common phenomenon. For example,
it happens when the theory has different branches, or loosely speaking
different phases. More concretely, when we consider the path integral
for a theory with a (discrete or continuous) moduli space of vacua,
this path integral is inherently multivalued, the ambiguity coming
from the ambiguity in the choice of vacuum. It is also well-know that,
in some cases, the different branches can be continuously related to
each other by analytic continuation
\cite{ferQPS1,ferQPS2,phase1,phase2}. We are going to explain that the
multivaluedness of $\F(z)$ can be understood in similar terms. Note
however that there is an important difference: in cases like $\F(z)$,
for which $z$ enters the path integral only through the insertion of a
local operator, the boundary conditions at infinity cannot depend on
$z$. This means that the gauge theory vacuum \eqref{cavacua} must be
the same for all branches of $\F$.

The characteristic function can be expressed, \emph{in a manifesly
gauge invariant way for all values of $z$}, as an integral over
anticommuting variables,
\be\label{cfgi} \det(z-M) =z^{N}\int\!\d\psi\d\chi\,
\exp\Bigl[\chi_{a}\bigl(\delta^{a}_{b} - z^{-1}M^{a}_{\
b}\bigr)\psi^{b}\Bigr]\, .\ee
Strictly speaking, this formula is only valid classically. Quantum
mechanically, a smearing in space-time is necessary, and the integral
in the right hand side of \eqref{cfgi} is to be interpreted as being a
\emph{path} integral localized around a certain space-time point $x$,
in the limit where the smearing distribution goes to the Dirac
$\delta$. This smearing is very similar in nature to the smearing of
the glueball field introduced in \cite{AIVW}. In particular,
$(\chi_{a}M^{a}_{\ b}\psi^{b})^{k}$ can be non-zero for arbitrary
powers $k$. When expanded around $\psi^{a}=\chi_{a}=0$, the integral
\eqref{cfgi} then yields an infinite series in $z$. Using Wick's
theorem and the contraction $\smash{\wick[u]{1}{<1\chi_{a}>1\psi^{b}}}=
\delta_{a}^{b}$ in the limit where the smearing is removed, it goes to
the expansion \eqref{detexp}.

Expanding around $\psi^{a}=\chi_{a}=0$ is not the only possibility.
Indeed, the classical equations of motion for \eqref{cfgi} read
\be\label{clchf}
M^{a}_{\ b}\psi^{b} = z\psi^{a}\, ,\quad \chi_{a}M^{a}_{\ b}
= z\chi_{b}\, .\ee
When $z$ coincides with an eigenvalue of $M$, there is a moduli space
of solutions (``Higgs branch'') parametrized by the gauge invariant
``meson''
\be\label{sigmadef}\sigma = \chi_{a}\psi^{a}\, .\ee
This is a hint that \eqref{cfgi} can indeed be multivalued. Quantum
mechanically, the eigenvalues of $M$ fluctuate. On the one hand,
these fluctuations can lift the flat direction in $\sigma$, and on the
other hand the ``Higgs branch'' can exist for any $z$, with a given
value of $\langle\sigma\rangle = \sigma_{\text H}(z)$. This behaviour
is reproduced for example by a simple toy model for which the quantum
corrections are modeled by a gaussian integration over $M$ (note
however that the gauge theory fluctuations are much more subtle:
single trace operators behave as if the eigenvalues were fluctuating,
whereas multi-trace correlators factorize as if they were dominated by
a non-fluctuating master field).

Taking the derivative with respect to $z$, the formula \eqref{cfgi}
shows that the condensate $\langle\sigma\rangle$ is nothing but the
gauge theory resolvent $\R$,
\be\label{Rsigma}
\langle\sigma\rangle (z) = \R(z)\, .\ee
We thus have a nice interpretation of the multivaluedness of $\R(z)$,
that correspond to the different ``phases'' of the integral
\eqref{cfgi}.

\subsubsection{A simple proof of the quantization conditions}

The above interpretation may be innocent-looking, but it has deep
consequences. The path integral \eqref{cfgi} can be multivalued
because it can have several ``vacua.'' In particular, once we pick a
vacuum, all the ambiguity must be lifted. But the vacua are
characterized by the set of gauge invariant expectation values that
can be built from the elementary fields over which we integrate. In
the case of \eqref{cfgi}, all we have is the gauge invariant meson
\eqref{sigmadef}, obtained by taking the derivative with respect to
$z$, or higher powers of $\sigma$, obtained by taking higher
derivatives. This implies that, even though \eqref{cfgi} may be a
multivalued function of $z$, it must have precisely the same sheets
and branch cuts as $\langle\sigma\rangle (z)=\R(z)$. In other words,
it must be a meromorphic function on the Riemann surface \eqref{RS}.

Let us analyze the consequences of this fact, focusing on the case of
the theory \eqref{Lagdef} for concreteness. By integrating the
relation \eqref{RFrel} from the point at infinity on the classical
sheet to $z$, we obtain
\be\label{Fintegral}\F(z) =
\mu_{0}^{N}\exp\int_{\mu_{0}}^{z}\!\R(z')\,\d z'\, .\ee
This formula shows that $\F$ is generically a multivalued function on
the curve \eqref{RS}, because one must specify the contour from the
point at infinity $\mu_{0}$ to $z$ to do the integral. A first
simplification occurs thanks to the conditions \eqref{genNI}. We can
define unambiguously $\F(z)$ on the classical sheet by integrating
along any path that remains on this sheet. The answer does not depend
on how many times the path circles around the branch cuts of $\R$.
This means that $\R$ and $\F$ have the same branch cuts on the
classical sheet, and the latter can be equally well characterized by
the properties \eqref{psasymp} or by
\be\label{asymp2}\F(z)\underset{z\rightarrow\infty}{\sim}z^{N} \,
.\ee
We can say more about the classical sheet branch cuts of $\F$: as for
$\R$, they are square root branch cuts. This is so because going twice
across the cut corresponds to going along a contour whose homotopy
class is trivial. Let us also note that the poles of $\R$ become
simple zeros of $\F$. Complications arise when we consider the
analytic continuations across the classical sheet branch cuts. Unlike
for $\R$, if we go through two different branch cuts, we obtain a
priori two inequivalent continuations of $\F$. Equation
\eqref{Fintegral} shows that this does not occur if and only if
\be\label{expqc}\exp\oint_{\gamma_{IJ}}\!\!\R(z)\,\d z =1\, ,\ee
in which case $\F$ and $\R$ are defined on the same surface \eqref{RS}
as required. Obviously, \eqref{expqc} is equivalent to the
quantization conditions \eqref{genqc}.

We are now going to show that the same result also follows from a
rigorous implementation of the constraints \eqref{permute} on $\F$.

\subsection{The permutation symmetry}
\label{QCsec}
\subsubsection{Dependence of the eigenvalues}

We now focus on the model \eqref{Lagdef} with $M=X$ for concreteness.
Let us go to a gauge where $X$ is diagonal (this is the supersymmetric
version of 't~Hooft's abelian projection \cite{thooftAP}),
\be\label{Xdiag} X = \diag (x_{1},\ldots,x_{N})\, .\ee
The gauge fixing procedure in the gauge theory, which mimics the gauge
fixing in the associated matrix model \eqref{MMdef}, was discussed in
\cite{kaza}. The eigenvalues $x_{I}$ are operators in the quantum
theory. In a given vacuum \eqref{cavacua}, we expand each eigenvalue
around its classical value. For example, we may choose
\be\label{eigenexp} x_{i} = a_{I} + \hat x_{i}\quad \text{for}\ 
\sum_{J<I}N_{J}+1\leq i\leq \sum_{J\leq I}N_{J}\, . \ee
The gauge theory resolvent can be decomposed in a natural way as
\be\label{Rdecomp}\R(z) = \sum_{I=1}^{\tilde d}N_{I}\R_{I}(z)+
\sum_{Q=1}^{\ell}\frac{\nu_{Q}}{z-b_{Q}}\ee
with
\be\label{RIdef} \R_{I}(z) =\frac{1}{N_{I}} \oint_{\alpha_{I}}\frac{\d
z'}{2i\pi}\,\frac{\R(z')}{z-z'} =
\Bigl\langle\frac{1}{z-x_{i}}\Bigr\rangle
\, \cvp\quad \sum_{J<I}N_{J}+1\leq i\leq \sum_{J\leq I}N_{J}\, .\ee
The variable $z$ in \eqref{RIdef} is taken to be outside the contour
$\alpha_{I}$ on the classical sheet (see Figure \ref{fig1}). Using
\eqref{genNI}, we see that the function $\R_{I}$ has only one square
root branch cut $[a_{I}^{-},a_{I}^{+}]$ on the classical sheet. The
decomposition \eqref{Rdecomp} yields a similar decomposition of the
characteristic function,
\be\label{Fdecomp} \F(z) = \prod_{I=1}^{\tilde
d}\F_{I}(z)^{N_{I}}\prod_{Q=1}^{\ell}(z-b_{Q})^{\nu_{Q}}\, ,\ee
with
\be\label{FIdef}\F_{I}(z) =
\mu_{0}\exp\int_{\mu_{0}}^{z}\!\R_{I}(z')\,\d z'\, .\ee
Equation \eqref{Fdecomp} is the quantum version of the
classical formula
\be\label{Fcl}\F_{\text{cl}}(z) = \prod_{I=1}^{\tilde
d}(z-a_{I})^{N_{I}}\prod_{Q=1}^{\ell}(z-b_{Q})^{\nu_{Q}}\, .\ee

The basic property of the expansion \eqref{Rdecomp} is that each
factor $\R_{I}$ depends only on the eigenvalues $x_{i}$ for
$\sum_{J<I}N_{J}+1\leq i\leq \sum_{J\leq I}N_{J}$. The same property
is true for the $\F_{I}$ that are expressed by \eqref{FIdef} in terms
of $\R_{I}$ only. More precisely, in the quantum theory, the classical
eigenvalue $x_{i}$ is replaced by the full set of correlators
$\{\langle x_{i}^{k}\rangle, k\geq 1\}$, or equivalently by a
probability distribution $\rho_{i}$. This probability distribution is
the same for all $i$ satisfying the condition $\sum_{J<I}N_{J}+1\leq
i\leq \sum_{J\leq I}N_{J}$, and will be denoted by $\rho_{I}$. The
resolvent $\R_{I}$ depends only on $\rho_{I}$ and can be expressed as
\be\label{RirhoI} \R_{I}(z) = \R(z;\rho_{I}) =
\int\frac{\rho_{I}(z')\,\d z'}{z-z'}\,\cdotp\ee
Conversely, the distribution $\rho_{I}$ is given by the discontinuity
across the classical sheet branch cut,
\be\label{rhoIR}\rho_{I}(x) = \frac{i}{2\pi}\bigl( \R_{I}(x+i\epsilon)
- \R_{I}(x-i\epsilon) \bigr)\, ,\quad x\in [a_{I}^{-},a_{I}^{+}]\,
.\ee

A potentially confusing point\footnote{I thank the referee for raising
this point.} is that the expansions \eqref{Req} and \eqref{Rdecomp} of
$\R$ are similar-looking (at least for $\Nf=0$), and that the forms
$\R_{I}\d z$ and $\psi_{I}\d z$ both satisfy
\be\label{psinorm22}
\frac{1}{2i\pi}\oint_{\alpha_{J}}\psi_{I}\,\d z = \delta_{IJ}\, ,\quad
\frac{1}{2i\pi}\oint_{\alpha_{J}}\R_{I}\,\d z = \delta_{IJ}\, .\ee
These similarities are very misleading, and the expansions \eqref{Req}
and \eqref{Rdecomp} are of \emph{completely different origins}. In
particular, the $\R_{I}$ defined by \eqref{RIdef} are not
single-valued on the curve \eqref{RS} (even though the sum
\eqref{Rdecomp} obviously is), they have only one branch cut
$[a_{I}^{-},a_{I}^{+}]$ on the classical sheet (the
$[a_{J}^{-},a_{J}^{+}]$ for $J\not =I$ are not branch cuts), and they
have in general an \emph{infinite} number of sheets (they are not
algebraic functions). This is very different from the $\psi_{I}$, that
are single-valued on \eqref{RS}, have branch cuts on
$[a_{J}^{-},a_{J}^{+}]$ for all $J$ on the classical sheet, and
satisfy a degree two algebraic equation. Moreover, for the forms
$\psi_{I}\d z$, the constraint \eqref{psinorm22} is the non-trivial
statement of Poincar\'e duality. In particular, the fact that the
$\alpha_{J}$-period integral of $\psi_{I}\d z$ is zero for $J\not = I$
is a non-trivial statement (it gives constraints on the polynomials
$p_{I}$ in \eqref{psiIform}). On the other hand, the similar statement
on $\R_{I}\d z$ is \emph{trivial}. It just follows from the fact that
$\R_{I}$ has no branch cut or other singularity on
$[a_{J}^{-},a_{J}^{+}]$ or inside $\alpha_{J}$, and thus the contour
can be deformed to a point when doing the corresponding integral.

The functions $\R_{I}$ or $\F_{I}$ are important for our purposes
because they depend only on $\rho_{I}$, and thus the action of the
permutation of the eigenvalues can be easily studied on them as we
shall see below. This simple property is not shared by the $\psi_{I}$.

\subsubsection{Permuting the eigenvalues}

In the sake of simplicity, let us first consider the case where all
the $N_{I}=1$ and $\nu_{Q}=0$ (this is relevant, for example, to the
solution of the $\nn=2$ theories, see Section 5). The generalized
Konishi anomalies imply that the distributions $\rho_{I}$ are
parametrized by the branching points $a_{K}^{\pm}$ of the curve
\eqref{RS} and the parameters $b_{Q}$ (the latter enters the solution
through the disk resolvent in \eqref{DVpot}),
\be\label{FIdep}\F_{I}(z) = \F(z;\rho_{I})=
\F_{I}(z;a_{K}^{\pm};b_{Q})\, .\ee
The permutation operator $\sigma_{IJ}=\sigma_{JI}$ exchanges the
$I^{\text{th}}$ and $J^{\text{th}}$ eigenvalues, that is to say
exchanges the distributions $\rho_{I}$ and $\rho_{J}$. In explicit
formulas, it is straightforward to see from the solution described in
Section 2 that this amounts to exchanging the branching points
$a_{I}^{\pm}$ and $a_{J}^{\pm}$, because we have
\be\label{rhoperm}
\begin{split}
\rho_{I}(x;\ldots,a_{I}^{\pm},\ldots,a_{J}^{\pm},\ldots;b_{Q})& =
(\sigma_{IJ}\cdot
\rho_{J})(x;\ldots,a_{I}^{\pm},\ldots,a_{J}^{\pm},\ldots;b_{Q})\\
&=\rho_{J}(x;\ldots,a_{J}^{\pm},\ldots,a_{I}^{\pm},\ldots;b_{Q})\, .
\end{split}\ee
In particular, this shows that
\be\label{exc1}
\begin{split}
\F_{I}(z;\ldots,a_{I}^{\pm},\ldots,a_{J}^{\pm},\ldots;b_{Q})& =
(\sigma_{IJ}\cdot\F_{J})(z;\ldots,a_{I}^{\pm},\ldots,a_{J}^{\pm},
\ldots;b_{Q})\\
&= \F_{J}(z;\ldots,a_{J}^{\pm},\ldots,a_{I}^{\pm},\ldots;b_{Q})\, ,
\end{split}\ee
and
\be\label{trivp}
\begin{split}
\F_{K}(z;\ldots,a_{I}^{\pm},\ldots,a_{J}^{\pm},\ldots;b_{Q})& =
(\sigma_{IJ}\cdot\F_{K})(z;\ldots,a_{I}^{\pm},\ldots,a_{J}^{\pm},
\ldots;b_{Q})\\
&= \F_{K}(z;\ldots,a_{J}^{\pm},\ldots,a_{I}^{\pm},\ldots;b_{Q})
\quad\text{for}\ K\not\in\{I,J\}\, .
\end{split}\ee
These equations are of course valid on all sheets. If $\F_{I,0}(z)$ is
the value of $\F_{I}(z)$ on the classical sheet, and $\F_{I,1}(z)$ the
value on the second sheet that is glued to the classical sheet by the
branch cut $[a_{I}^{-},a_{I}^{+}]$ ($\F_{I}$ has in general additional
sheets, but they are irrelevant for our purposes), then
\begin{align}\label{pons1}
\F_{I,0}(z;\ldots,a_{I}^{\pm},\ldots,a_{J}^{\pm},\ldots;b_{Q})& =
\F_{J,0}(z;\ldots,a_{J}^{\pm},\ldots,a_{I}^{\pm},\ldots;b_{Q})\, ,\\
\label{pons2}
\F_{I,1}(z;\ldots,a_{I}^{\pm},\ldots,a_{J}^{\pm},\ldots;b_{Q})& =
\F_{J,1}(z;\ldots,a_{J}^{\pm},\ldots,a_{I}^{\pm},\ldots;b_{Q})\, .
\end{align}
We can now implement the constraint \eqref{permute} on $\F$. It is
equivalent to the identity
\be\label{FCexp}\begin{split}
\F(z;\ldots,a_{I}^{\pm},\ldots,a_{J}^{\pm},\ldots;b_{Q})&=
(\sigma_{IJ}\cdot\F)(z;\ldots,a_{I}^{\pm},\ldots,a_{J}^{\pm},
\ldots;b_{Q})\\
&= \F(z;\ldots,a_{J}^{\pm},\ldots,a_{I}^{\pm},\ldots;b_{Q})\, ,
\end{split}\ee
which must be valid on all sheets.\footnote{Let us stress that the
action of $\sigma_{IJ}$ does not mix up the different sheets. The
$a^{\text{th}}$ sheet can be characterized by the coefficients
$c_{a,k}(\rho_{I})$ of the large $z$ expansion, which have a certain
sheet-dependent functional dependence on the distribution $\rho_{I}$.
Permuting $\rho_{I}$ and $\rho_{J}$ clearly does not change $a$.} On
the classical sheet, this is of course automatically satisfied by
using \eqref{pons1},
\be\label{cp1}
\begin{split}
\F_{0}(z;\ldots,a_{I}^{\pm},\ldots,a_{J}^{\pm},\ldots;b_{Q}) & =
\prod_{K=1}^{N}\F_{K,0}(z;\ldots,a_{I}^{\pm},\ldots,a_{J}^{\pm},
\ldots;b_{Q})\\
& = \F_{0}(z;\ldots,a_{J}^{\pm},\ldots,a_{I}^{\pm},\ldots;b_{Q})\, .
\end{split}\ee
Let us now consider the analytic continuation $\F_{1}^{(I)}$ of
$\F_{0}$ through the $I^{\text{th}}$ branch cut (a priori,
$\F_{1}^{(I)}$ and $\F_{1}^{(J)}$ could be distinct functions if
$I\not = J$). Remembering that $[a_{I}^{-},a_{I}^{+}]$ is not a branch
cut of the $\F_{K,0}$ for $K\not = I$, we find
\be\label{F1Idef}\F_{1}^{(I)}(z) = \F_{I,1}(z) \prod_{K\not =
I}\F_{K,0}(z)\, .\ee
The identity \eqref{FCexp} then yields
\begin{multline}\label{masterid}
\F_{I,1}(z;\ldots,a_{I}^{\pm},\ldots,a_{J}^{\pm},\ldots;b_{Q})
\prod_{K\not = I}\F_{K,0}
(z;\ldots,a_{I}^{\pm},\ldots,a_{J}^{\pm},\ldots;b_{Q}) =\\
\F_{I,1}(z;\ldots,a_{J}^{\pm},\ldots,a_{I}^{\pm},\ldots;b_{Q})
\prod_{K\not = I}\F_{K,0}
(z;\ldots,a_{J}^{\pm},\ldots,a_{I}^{\pm},\ldots;b_{Q})\, .
\end{multline}
Using \eqref{pons1} and \eqref{pons2} on the right hand side of this
equation, we obtain
\begin{multline}\label{masterid2} 
\F_{I,1}(z;\ldots,a_{I}^{\pm},\ldots,a_{J}^{\pm},\ldots;b_{Q})
\F_{J,0}(z;\ldots,a_{I}^{\pm},\ldots,a_{J}^{\pm},\ldots;b_{Q}) =\\
\F_{J,1}(z;\ldots,a_{I}^{\pm},\ldots,a_{J}^{\pm},\ldots;b_{Q})
\F_{I,0}(z;\ldots,a_{I}^{\pm},\ldots,a_{J}^{\pm},\ldots;b_{Q})\, .
\end{multline}
This identity simply tells us that the analytic continuation of
$\F_{0}$ through the $J^{\text{th}}$ branch cut,
\be\label{F1Jdef}\F_{1}^{(J)}(z) = \F_{J,1}(z) \prod_{K\not =
J}\F_{K,0}(z)\, ,\ee
must be equal to the analytic continuation \eqref{F1Idef} through the
$I^{\text{th}}$ branch cut. Using \eqref{Fintegral}, this immediately
yields the quantization condition \eqref{genqc}.

The most general case or arbitrary $N_{I}$ and $\nu_{Q}$ can be
treated along similar lines. One must take into account that all the
eigenvalues that are sitting at the same point classically are
constrained by the generalized Konishi equations to have the same
resolvent \eqref{RIdef} quantum mechanically. Moreover, the
eigenvalues at $b_{Q}$ cannot fluctuate. The constraints on $\F$ are
then obtained by exchanging the $N_{I}$ eigenvalues at $a_{I}$ with
the $N_{J}$ eigenvalues at $a_{J}$. This is a symmetry as long as one
exchanges $N_{I}$ and $N_{J}$ at the same time, corresponding to a
combination of permutations that exchange the $I^{\text{th}}$ and
$J^{\text{th}}$ blocks of the matrix $X$.

\section{One more Ward identity}
\setcounter{equation}{0}

Taking into account \eqref{DVpotder}, the quantization conditions
\eqref{genqc} are equivalent to
\be\label{DVsmall} \frac{\partial\wdv}{\partial s_{I}} =
\frac{\partial\wdv}{\partial s_{J}}\, \cdotp\ee
We can use these $\tilde d -1$ equations to express the $\tilde d$
parameters $s_{I}$ in terms of a single parameter $s = \sum_{I}s_{I}$,
\be\label{sIhat} s_{I}= \hat s_{I}(s)\, .\ee
This parameter $s$ is the last indeterminate in our problem. The
equation \eqref{sumsI} shows that it is identified with the glueball
superfield \eqref{gluedef} expectation value, $s=\langle S\rangle$. It
is natural to introduce a new analytic function defined by
\be\label{DVglue}\whdv (s) = \wdv\bigl(\hat s_{I}(s)\bigr)\, .\ee
To complete the proof of the Dijkgraaf-Vafa conjecture
\eqref{DVqem}, we need to show that
\be\label{DVlast}\whdv'\bigl(s=\langle S\rangle\bigr) = 0\, ,\ee
where the $'$ denotes the derivative with respect to $s$. Let us note
that when $\tilde d = 1$ and the low energy gauge group has a single
$\text{U}(N_{1})$ factor, $\whdv=\wdv$ and equation \eqref{DVlast} is
all we have. This simple ``one-cut'' case contains already a great
deal of interesting non-perturbative physics \cite{fer0,ferQPS1}. The
aim of this Section is to show that \eqref{DVlast} follows from a
simple Ward identity.

\subsection{The quantum glueball superpotential}

The simplest way to present the proof is to use the so-called quantum
glueball superpotential $W(s)$. This is defined to be the Legendre
transform of $\wl$, defined in \eqref{wlowdef}, with respect to the
gauge coupling constant. More precisely, in the theory with $\Nf$
flavors, the expectation value of the glueball superfield
\eqref{gluedef} is given by the equation \eqref{wlder2bis} (with the
trivial modification $\ln\La^{2N-\Nf}\rightarrow 2i\pi\tau$ when
$\Nf=2N$). This equation can be inverted to find $\ln\La^{2N-\Nf}$ as
a function $f$ of $\langle S\rangle = s$. Indicating the dependence on
the various couplings explicitly, with in particular \eqref{tKdef},
the glueball superpotential is then defined by
\be\label{Wgluedef} W(s;t_{K},\ln\La^{2N-\Nf}) =
\wl\bigl(t_{K},f(s;t_{K})\bigr) + \bigl(\ln\La^{2N-\Nf} -
f(s;t_{K})\bigr) s \, . \ee
By construction, $W(s)$ is linear in $\ln\La^{2N-\Nf}$, and satisfies
the equations
\begin{align}\label{ons1} &\frac{\partial W}{\partial g_{k}}=
\frac{1}{k+1}\bigl\langle \Trf X^{k+1}\bigr\rangle\, ,\quad
\frac{\partial W}{\partial m_{f,k}^{\ f'}} = \bigl\langle 
{}^{T}\tilde Q^{f}X^{k}Q_{f'}\bigr\rangle\, ,\\
\label{ons3} & W'\bigl(s=\langle S\rangle\bigr) = 0\, .\end{align}
These equations are immediate consequences of the definition
\eqref{Wgluedef} and of \eqref{wlder2} and \eqref{wlder2bis}. In
particular, \eqref{ons3} implies that if we could prove that
\be\label{eqtoder} W \overset{?}= \whdv\, ,\ee
then the last Dijkgraaf-Vafa equation \eqref{DVlast} would automatically
follow. We are thus going to derive \eqref{eqtoder} in the following.

This derivation will necessitate a little bit of more work, but from
what we already know we can immediately deduce that
\be\label{wwdvrel} W(s;t_{K},\ln\La^{2N-\Nf}) =
\whdv(s;t_{K},\ln\La^{2N-\Nf}) + h(s)\, ,\ee
where $h$ depends on $s$ only (not on the couplings $t_{K}$ or $\La$)
and $h(0)=0$. This last condition on $h$ follows from the fact that,
by construction, $W$ and $\whdv$ have the same classical,
$s\rightarrow 0$, limit. Moreover, the dependence on $\ln\La^{2N-\Nf}$
is through the simple linear term $s\ln\La^{2N-\Nf}$ on both sides of
\eqref{wwdvrel}. This is clear for $W$ from the definition
\eqref{Wgluedef}, and it is also true for $\whdv$, that depends on
$\La$ only through the normalization factor \eqref{Nc1} in the matrix
integral. Actually, the $\La$-dependence of $\mathscr N$ was chosen
precisely for this matching to be valid. Finally, \eqref{DVsmall},
\eqref{sIhat} and \eqref{DVglue} imply that $\partial_{t_{K}}\wdv =
\partial_{t_{K}}\whdv$. Then the equations \eqref{ons1} on the one
hand, and \eqref{weakDV} on the other hand, show that the
$t_{K}$-dependence on both sides of \eqref{wwdvrel} must be the same.

\subsection{The last Dijkgraaf-Vafa equation}

Thus we have to prove that $h=0$, or more precisely that $h$ must be a
linear function, since a linear shift with a coupling-independent
slope can be absorbed in the normalization $\mathscr N$ (or
equivalently in a rescaling of $\La$ on the matrix model side).

Our strategy is as follows. Let us introduce the first order
differential operator
\be\label{diff}\mathcal D = s\frac{\partial}{\partial s} +
\sum_{k}g_{k}\frac{\partial}{\partial g_{k}}\, \cdotp\ee
We are first going to show that matrix model identities imply that
\be\label{dwdv} \mathcal D\cdot \whdv = \whdv\, .\ee
This equation is just an identity that follows from the definition of 
$\whdv$. It strongly depends in particular on the $s$-dependence of
the normalization factor $\mathscr N$ \eqref{Nc1}. Put in another way,
we have chosen this $s$-dependence precisely for the equation
\eqref{dwdv} to be valid. In particular, it fixes uniquely the
Veneziano-Yankielowicz contributions to $\whdv$. Second, we are going 
to show that a gauge theory Ward identity implies that
\be\label{dW}\mathcal D\cdot W = W\, .\ee
Combining \eqref{dwdv}, \eqref{dW} and \eqref{wwdvrel}, we
obtain 
\be\label{heq} \mathcal D\cdot h = h = sh'\, ,\ee
whose most general solution is a linear function, as announced.

\subsubsection{Matrix model identities}

The simplest way to derive \eqref{dwdv} is to start from the matrix
integral and to perform the infinitesimal variations
\be\label{varMMid} \delta s_{I} = \epsilon s_{I}\, ,\quad \delta s =
\epsilon s\, ,\ee
on both sides of \eqref{MMdef}. The matrix model vacuum
\eqref{mmvacua} is left invariant, because \eqref{ff} implies that the
$n_{I}$ are left invariant. By taking into account the $s$-dependence
of the normalization factor $\mathscr N$ given by \eqref{Nc1}, we
obtain immediately
\be\label{idmm0}
\begin{split}
-2\Fb + \sum_{I}s_{I}\frac{\partial\Fb}{\partial s_{I}}&
= \frac{s}{n}\bigllangle \Trf\wt(X)+{}^{T}\tilde Q^{f}m_{f}^{\
f'}(X)Q_{f'}\bigrrangle -\frac{s^{2}\Nf}{n}\\
& = -\sum_{k}g_{k}\frac{\partial\Fb}{\partial g_{k}} + \frac{s}{n}
\bigllangle {}^{T}\tilde Q^{f}m_{f}^{\ f'}(X)Q_{f'}\bigrrangle
-\frac{s^{2}\Nf}{n}\, \cdotp \end{split}\ee
Expanding in powers of $1/n$ as in \eqref{Flargen}, and using
\be\label{mmloopsei} 
\bigllangle {}^{T}\tilde Q^{f}m_{f}^{\
f'}(X)Q_{f'}\bigrrangle_{\text{disk}} = sN_{f}\ee
which is a direct consequence of the matrix model loop equations (see
for instance \cite{seifla}), we obtain
\begin{align}\label{idmm1} 
\sum_{I}s_{I}\frac{\partial\Fp}{\partial s_{I}}  +
\sum_{k}g_{k}\frac{\partial\Fp}{\partial g_{k}} & = 2\Fp\, ,\\
\label{idmm2}
\sum_{I}s_{I}\frac{\partial\Fd}{\partial s_{I}}
+\sum_{k}g_{k}\frac{\partial\Fd}{\partial g_{k}} &= \Fd\, .\end{align}
Relations similar to \eqref{idmm1} have been considered in special
cases \cite{matone2}. Combining \eqref{idmm1} and \eqref{idmm2} with
\eqref{DVpot}, we get
\be\label{idmm3} \sum_{I}s_{I}\frac{\partial\wdv}{\partial s_{I}}
+\sum_{k} g_{k}\frac{\partial\wdv}{\partial g_{k}} = \wdv\, ,\ee
and equation \eqref{dwdv} then follows from the definition
\eqref{DVglue}.

\subsubsection{Gauge theory identities}

The anomaly equations discussed in Section 2.3 are generalizations of
more standard (anomalous) Ward identities associated with the
symmetries of the gauge theory \eqref{Lagdef}. There is however a
useful non-anomalous $\u_{\text R}$ symmetry that yields a new
constraint. If we denote by $\theta$ the superspace coordinates, by
$\lambda$ the gluino, and $w$ any effective superpotential, the
assignement of charges under this $\u_{\text R}$ is as follows:

\be\label{asign}
\begin{matrix}
& \theta & \lambda  & X & Q_{f} & \tilde Q^{f} &  
g_{k} & m_{f,q}^{\ f'} & w \\
\u_{\text R} & 1 & 1 & 0 & 1 & 1 & 2 & 0 & \hphantom{,\,} 2\, .
\end{matrix}\ee

This symmetry implies that
\be\label{wlWard} \wl = \sum_{k}g_{k}\frac{\partial\wl}{\partial g_{k}}
= \bigl\langle\Trf\wt(X)\bigr\rangle\, .\ee
After Legendre transforming to $W$, we obtain \eqref{dW}, which
completes the proof.

\subsubsection{The theory with two flavors}

The theory with two flavors can be treated along similar lines. 

On the matrix model side, if we denote collectively by $t_{K}$ the
couplings in the tree-level superpotential \eqref{wtree2ad}, the
planar partition function defined by \eqref{twoMM} with the
normalization factor $\mathscr N$ given by \eqref{norm2MM} satisfies
\be\label{id2mm1} \sum_{I}s_{I}\frac{\partial\Fp}{\partial s_{I}} +
\sum_{K}t_{K}\frac{\partial\Fp}{\partial t_{K}}+\frac{s^{2}}{2} =
2\Fp\, . \ee
This is the analogue of \eqref{idmm1} for the model with fundamental
flavors. If
\be\label{Ddef2mm}\mathcal D = s\frac{\partial}{\partial s} +
\sum_{K}t_{K}\frac{\partial}{\partial t_{K}}\, \cvp\ee
equation \eqref{id2mm1} yields
\be\label{id2mm2} \mathcal D\cdot\whdv - Ns = \whdv\ee
for the Dijkgraaf-Vafa superpotential defined by \eqref{wdv2mm} and
\eqref{DVglue}.

On the gauge theory side, we use the anomalous $\u_{\text R}$ symmetry

\be\label{asign2}
\begin{matrix}
& \theta & \lambda  & X & Y & \La^{N} &  t_{K} & w \\
\u_{\text R} & 1 & 1 & 0 & 0 & -2N & 2 & \hphantom{,\,} 2\, .
\end{matrix}\ee

\noindent to show that
\be\label{wlWard2} \wl = \sum_{K}t_{K}\frac{\partial\wl}{\partial t_{K}}
- N\frac{\partial\wl}{\partial\ln\La^{N}} =
\bigl\langle\Trf\wt(W,Y)\bigr\rangle - N\bigl\langle S\bigr\rangle\,
.\ee
Legendre transforming, we obtain
\be\label{dW2mm} \mathcal D\cdot W -Ns = W\, .\ee
Comparing this with \eqref{id2mm2}, and using the analogue of
\eqref{wwdvrel}, we conclude that $W=\whdv$.

\subsection{A note on the most general solutions to the constraints}

There is one implicit assumption that we have made so far.\footnote{I
thank the referee for raising this point.} If $N_{I}\not = 0$, we have
taken for granted that the branch cut $[a_{I}^{-},a_{I}^{+}]$ opens
up, or equivalently that $s_{I}$ is not identically zero. This is
necessary to obtain the full set of quantization conditions
\eqref{genqc}. However, another possibility is to let some of the
$s_{I}$ to be identically zero (let's say for $1\leq I\leq \delta$),
even though the corresponding unbroken factors of the gauge group
$\text{U}(N_{I})$ are non-trivial. The remaining parameters $s_{K}$
for $K\geq \delta +1$ are then fully determined by the quantization
conditions $\oint_{\gamma_{KL}}\R(z)\d z\in 2i\pi\mathbb Z$ for
$K,L>\delta$ that follow from the analysis of Section 3 as well as
from the additional constraint discussed above (for example
\eqref{wlWard}). Let us emphasize that the solutions so obtained are
different from the $N_{I}=0$ solutions (which also have $s_{I}=0$),
because the $\alpha_{I}$-period of $\R(z)$ are constrained to be equal
to $N_{I}$.

What are we supposed to do with these additional solutions? One
consistent possibility is to consider that they are unphysical and
disregard them (this is consistent because they cannot be obtained by
analytic continuation from the other solutions: they belong to a
different phase). It is however natural to try to interpret them,
since they are good solutions of our set of constraints. Clearly they
cannot correspond to the usual vacua \eqref{cavacua}, and thus if they
are physical they imply the existence of new vacua, that we denote for
reasons to become clear in a moment
\be\label{mgv}
|N_{1},\text{KS};\ldots;N_{\delta},\text{KS};N_{\delta+1},k_{\delta+1};
\ldots;N_{d},k_{d};\nu_{1};\ldots;\nu_{\ell}\rangle\, .\ee
For example, when the gauge group in unbroken, we would have, on top
of the usual vacua $|N,k\rangle$, an additional vacuum
$|N,\text{KS}\rangle$. In this vacuum, the expectation value of the
glueball superfield is zero by definition,
\be\label{KSvacua} \langle N,\text{KS}|S|N,\text{KS}\rangle = 0\, .\ee
The state $|N,\text{KS}\rangle$ has all the required properties to be
identified with the chirally symmetric vacuum postulated by Kovner and
Shifman in \cite{KSvac}. The more general cases \eqref{mgv} are the
immediate generalizations of the Kovner-Shifman vacua for unbroken
$\text{U}(N_{1})\times\cdots\times\text{U}(N_{d})$ gauge group.

The existence of the Kovner-Shifman vacua is controversial and has
never been rigorously established. It is interesting to find that they
naturally fit in the framework of the present paper, and that the most
general consistent solution actually corresponds to the general states
of the form \eqref{mgv}. For this reason, and also because, to my
knowledge, a convincing argument that rules them out has not appeared,
we believe that these hypothetical new phases of $\nn=1$ gauge
theories deserve further investigations. Let us stress however that a
very plausible possibility is that there exists additional constraints
that restrict the physical vacuum states to be of the standard form
\eqref{cavacua}.

The original Kovner-Shifman vacua \cite{KSvac}, if they exist, must
belong to the strongly coupled regime of $\nn=1$ gauge theories, and
this makes their study particularly difficult. However, we want to
point out that some of our generalized KS vacua \eqref{mgv} can be
made arbitrarily weakly coupled by adjusting the parameters. This
occurs when the unbroken gauge group has only abelian factors. For
example, this is the case for a Coulomb branch with symmetry breaking
pattern $\uN\rightarrow\u^{N}$, which is also relevant to $\nn=2$
super Yang-Mills \cite{CV}. Thus in principle we may have KS vacua in
$\nn=2$ as well! The existence of weakly coupled KS vacua is
particularly puzzling, because explicit instanton calculations are
then reliable and seem to rule them out. There might remain the
possibility that some subtleties of the semiclassical path integral
have been overlooked.

\section{The confinement and mass gap properties}
\setcounter{equation}{0}

In this last Section, we are going to apply the above results to the
problem of the creation of a mass gap and of color confinement in
$\nn=1$ super Yang-Mills theory. The strategy we use to prove
confinement is to deform a $\nn=2$ theory by adding a tree-level
superpotential, as explained by Seiberg and Witten in \cite{SW}. Our
contribution is to waive the two main non-trivial assumptions on which
this famous reasoning is based.

The first assumption is that the solution of the $\nn=2$ theory given
in \cite{SW}, and its generalizations, are correct. This can now be
derived, using the results of Sections 2, 3 and 4, following the
approach proposed in \cite{CV}. Many examples have been studied in the
literature, for instance in \cite{SWfromMM}. However, the particularly
interesting case of the $\Nf=2N$ theory, for which the Yang-Mills
coupling does not run, doesn't seem to have been published. Moreover,
even for the cases with $\Nf<2N$, which can be obtained from the
$\Nf=2N$ theory by integrating out the quarks, the analysis presented
in the literature is often incomplete, because the last Dijkgraaf-Vafa
equation \eqref{DVlast} is usually not implemented. For these reasons,
and also because the result is instructive, we briefly present the
derivation for the $\Nf=2N$ theory in the next subsection.

The second and most important assumption is that the $\nn=1$ theory
has a mass gap. This is explained in Section 5.2, where we show how
this fundamental property can actually be derived by combining the
solutions of both the $\nn=2$ and $\nn=1$ theories obtained
previously.

\subsection{The $\nn=2$ theory with $\Nf=2N$}

We consider the theory \eqref{Lagdef} with a degree $d=N+1$ tree-level
superpotential pa\-ra\-me\-tri\-zed as
\be\label{WtreeN2}\wt'(x) = \epsilon P_{N}(x) =
\epsilon\prod_{I=1}^{N}(x-a_{I}),\ee
polynomials $m_{f}^{\ f'}$ of the form
\be\label{mffN2}m_{f}^{\ f'}(X) = (X-m_{f})\delta_{f}^{f'}\, ,\ee
and we focus on the vacuum $|N_{I}=1;\nu_{Q}=0\rangle$ which has
$\tilde d = d=N$ and gauge symmetry breaking $\uN\rightarrow\u^{N}$.
The Riemann surface \eqref{RS} takes the form
\be\label{SWc1} C:\ y^{2} = P_{N}^{2} - p_{N-1}(x)\,
,\ee
for a certain polynomial $p_{N-1}$ of degree $\deg p_{N-1}\leq N-1$.

The $N$ unknown coefficients in $p_{N-1}$ are determined by the $N$
quantum equations of motion \eqref{intpe} and \eqref{DVlast}. As we
have explained in details, \eqref{intpe} is equivalent to the fact
that the characteristic function $\F$ is a meromorphic function on the
curve \eqref{SWc1}. It has two poles of order $N$ at infinity on the
classical and second sheets, and $2N$ simple zeros at $z=m_{f}$ on the
second sheet. Moreover, equation \eqref{DVlast}, or equivalently one
of the equation \eqref{qem3} (with $\ell=2N$, $U_{0}=1$ and
$\La^{2N-\Nf}$ replaced by $e^{2i\pi\tau}$ since we are in the theory
with zero $\beta$ function) is equivalent to a simple asymptotic
condition on $\F$. If we note, as in Section 3, $\F_{0}(z)$ the value
of $\F$ on the classical sheet and $\F_{1}(z)$ the value on the second
sheet, then we have
\be\label{lasteq}
\lim_{\mu_{0}\rightarrow\infty}
\frac{\F_{1}(\mu_{0})}{\F_{0}(\mu_{0})}= e^{2i\pi\tau}=q\, .\ee
The above constraints can be straightforwardly solved (full details
are given in \cite{ferrev}). The result is that the curve \eqref{SWc1}
must take the form
\be\label{N2sol} y^{2} = \prod_{I=1}^{N}(z-a_{I})^{2} - p_{N-1}(z) = 
H_{N}(z)^{2} - \frac{4 q}{(1-q)^{2}}\prod_{f=1}^{2N}(z-m_{f})\, ,\ee
where $p_{N-1}$ and $H_{N}^{2}$ are determined uniquely be the
requirement that the second equality in \eqref{N2sol} must hold. The
characteristic function is given by
\be\label{FN2sol} \F(z) = \frac{1-q}{2}\bigl( H_{N}(z) + y(z)\bigr)\,
.\ee
From this, we can deduce in particular the quantum chiral ring
relations and the gauge theory resolvent $\R(z)$. It is sometimes
convenient to express the curve \eqref{N2sol} in terms of the
parameters $e_{I}$ defined by the relations
\be\label{clmodrel2} \langle\Trf X^{p}\rangle = \sum_{I=1}^{N}
e_{I}^{p}\, ,\quad 1\leq p\leq N\, .\ee
There is a great confusion in the literature when trying to express
the Seiberg-Witten curve in terms of these physical parameters.
However, knowing the quantum chiral ring it is a straightforward
calculation. Introducing the polynomial
\be\label{Endef} E_{N}(z) = \prod_{I=1}^{N}(z-e_{I})\ee
we find
\be\label{HNscafin} H_{N} = E_{N} +\frac{q}{(1-q)^{2}}
\Bigl[\frac{\prod_{f=1}^{2N}(z-m_{f})}{E_{N}}\Bigr]_{\mathrm{Pol.}}\,
,\ee
where we take the polynomial part of the rational function on the
right hand side of \eqref{HNscafin}.

The above solution does not depend on the parameter $\epsilon$ in
\eqref{WtreeN2}, and thus applies to the $\nn=2$ theory that is
obtained in the $\epsilon\rightarrow 0$ limit \cite{CV}. Strictly
speaking, we have obtained the full set of chiral operator vacuum
expectation values. The $\nn=2$ prepotential that governs the low
energy effective action can then be straightforwardly derived using
Matone's relation and its generalizations \cite{matone}. Matone's
relation itself is a consequence of superconformal Ward identities, as
shown by Howe and West in \cite{west}. The result is that the period
matrix of the curve \eqref{N2sol} yields the low energy $\u$ coupling
constants, as expected.

The theory with $\Nf=2N$ has a $\beta$ function equal to zero, and
thus it has long been conjectured to have a non-trivial S-duality
invariance. Accordingly, the original curve for gauge group
$\mathrm{SU}(2)$, derived in \cite{SW}, was presented in terms of
modular forms for a coupling $\tau_{\mathrm{eff}}$. It was soon
realized, however, that even though $\tau_{\mathrm{eff}}$ and the
microscopic gauge coupling $\tau$ were identical at the classical
level, they could not be equal in the quantum theory \cite{Doreydis}.
Several curves were subsequently proposed for any gauge group $\suN$,
all modular covariant with respect to some coupling
$\tau_{\mathrm{eff}}$ whose relation with $\tau$ has remained unclear
(see for example the discussion in \cite{AY}). Remarquably, our
solution \eqref{N2sol} has an unambiguous dependence in the UV
coupling $\tau$ that is not consistent with S-duality. This is in
sharp contrast with the $\nn=1^{*}$ theory, for which the same
approach yields results in beautiful agreement with S-duality
\cite{DV,Sduality}.

\subsection{The mass gap and confinement}

We start from the pure $\nn=2$ theory, whose curve
\be\label{pure2sol} y^{2} = \prod_{I=1}^{N}(z-a_{I})^{2} -
4\La^{2N}\ee
is the limit of \eqref{N2sol} when the quark masses go to infinity.
The limit of equation \eqref{HNscafin} implies in this case that
\be\label{clmodulirel}\langle\Trf X^{p}\rangle =
\sum_{I=1}^{N}a_{I}^{p}\, ,\quad 1\leq p\leq N\, .\ee
We can go to the $\suN$ theory and eliminate the irrelevant global
massless $\u$ by imposing the tracelessness condition
$\sum_{I}a_{I}=0$. For generic values of the moduli $a_{I}$, the low
energy theory is then a free abelian $\u^{N-1}$ theory. There are $N$
special points on the moduli space where the curve \eqref{pure2sol}
degenerates. The behaviour of the low energy $\u$ couplings near these
singularities shows that a dyon of magnetic charge one becomes
massless there. The electric charge of the dyon can be any integer
$0\leq k\leq N-1$, depending on which singular point we consider.

We now turn on the tree-level superpotential
\be\label{wtN2}\wt(X) = \frac{1}{2} m X^{2}\, .\ee
If $m$ is infinitesimal, the resulting theory can be seen as a small
deformation of the original $\nn=2$ theory. However, the $\nn=1$
theory does not have a moduli space. The moduli of the $\nn=2$ theory
must be ``frozen'' when the superpotential \eqref{wtN2} is turned on.
In \cite{SW}, Seiberg and Witten assumed that the moduli are frozen
precisely at the singular points on moduli space. \emph{This is
equivalent to assuming that the $\nn=1$ theory has a mass gap.} Since
this is an important point, let us present a simple proof.

Suppose that the $\nn=1$ theory obtained by deforming $\nn=2$ by
adding \eqref{wtN2} has a mass gap. Since the creation of a mass gap
is a low energy phenomenon, we can analyze the problem in the
framework of the low energy effective action. When $m=0$, we have a
free abelian theory. The only way such a theory can create a mass gap
is through the usual Higgs mechanism. But this requires charged fields
to be present in the effective action. This occurs only at the
singular points on the $\nn=2$ moduli space. Conversely, assume that
by turning on \eqref{wtN2} the moduli are frozen at the singular
points. It is then possible to analyze directly the effect of $\wt$ in
the effective action using symmetry arguments, and then to show that
the massless dyon indeed condenses \cite{SW}. Since the field that
condenses is magnetically charged, \emph{this also implies (oblique)
confinement.} Using analyticity, one can actually compute the dyon
condensate for any value of $m$ \cite{DS,ferQPS1} and show that it is
non-zero including in the limit $m\rightarrow\infty$ where we flow to
the pure $\nn=1$ gauge theory.

We are now reaching the finishing line. By using the solution of the
$\nn=1$ theory with the tree-level superpotential $\eqref{wtN2}$
described in Sections 2--4, we can compute exactly the correlators
$\langle\Trf X^{p}\rangle$. Using \eqref{clmodulirel}, it is
straightforward to check that, indeed, \emph{these values correspond
exactly to the singular points on the $\nn=2$ moduli space.} From what
we have said before, this demonstrates that a mass gap is created and
that confinement occurs in the $\nn=1$ super Yang-Mills theory.

\subsection*{Acknowledgements}

This work is supported in part by the belgian Institut
Interuniversitaire des Sciences Nucl\'eaires (convention 4.4505.86),
by the Interuniversity Attraction Poles Programme (Belgian Science
Policy) and by the European Commission FP6 programme
MRTN-CT-2004-005104 (in association with V.\ U.\ Brussels). The author
is on leave of absence from Centre National de la Recherche
Scientifique, Laboratoire de Physique Th\'eorique de l'\'Ecole Normale
Sup\'erieure, Paris, France.

\renewcommand{\thesection}{\Alph{section}}
\renewcommand{\thesubsection}{\arabic{subsection}}
\renewcommand{\theequation}{A.\arabic{equation}}
\setcounter{section}{0}
\end{document}